\documentclass[a4paper,11pt]{article}

\usepackage[a4paper]{geometry}

\usepackage{amsmath,amsthm,amssymb}
\usepackage{amsfonts}

\usepackage{graphicx}
\usepackage{authblk}
\usepackage{color}
\usepackage[font=small,labelfont=bf]{caption}
\usepackage[linktoc=all]{hyperref}
\usepackage[numbers,sort&compress]{natbib}
\usepackage{doi}

\usepackage{eprint}

\providecommand{\keywords}[1]{\textbf{\small \textit{keywords:}} {\small #1}}

\def\mathlette#1#2{{\mathchoice{\mbox{#1$\displaystyle #2$}}%
                               {\mbox{#1$\textstyle #2$}}%
                               {\mbox{#1$\scriptstyle #2$}}%
                               {\mbox{#1$\scriptscriptstyle #2$}}}}
\renewcommand{\vec}[1]{\mathlette{\boldmath}{#1}}

\DeclareMathOperator\erf{erf}

\newcommand{\sigA}{\sigma_A}
\newcommand{\sigB}{\sigma_B}

\newcommand{\cP}{\mathcal{P}}

\newcommand{\alphabet}{\mathcal{K}}				
\def\GF{\mathcal{GF}}						
\def\Gaussian{\mathcal{N}}					

\newcommand{\rawA}{\vec{z}_A}
\newcommand{\rawB}{\vec{z}_B}

\newcommand{\measXA}{\vec{x}_A}
\newcommand{\measXB}{\vec{x}_B}
\newcommand{\measYA}{\vec{y}_A}
\newcommand{\measYB}{\vec{y}_B}
\newcommand{\measZA}{\vec{z}_A}

\newcommand{\rawlA}{\vec{\check{z}}_A}
\newcommand{\rawlB}{\vec{\check{z}}_B}
\newcommand{\rawmA}{\vec{\hat{z}}_A}
\newcommand{\rawmB}{\vec{\hat{z}}_B}

\newcommand{\quant}{\mathcal{Q}}

\title{Information Reconciliation for Continuous-Variable Quantum Key Distribution using Non-Binary Low-Density Parity-Check Codes}

\author[1]{Christoph Pacher\thanks{\href{mailto: christoph.pacher@ait.ac.at}{christoph.pacher@ait.ac.at}}}
\author[2]{Jesus Martinez-Mateo}
\author[3]{J\"{o}rg Duhme}
\author[4,5]{Tobias Gehring}
\author[6]{Fabian Furrer}
\affil[1]{\small Digital Safety \& Security Department, AIT Austrian Institute of Technology GmbH, Donau-City-Stra{\ss}e 1, 1220 Vienna, Austria}
\affil[2]{\small Center for Computational Simulation, Universidad Polit\'{e}cnica de Madrid,\authorcr
Campus de Montegancedo, 28660 Boadilla del Monte, Madrid, Spain}
\affil[3]{\small Institute of Theoretical Physics, Leibniz Universit\"at Hannover,\authorcr
Appelstra\ss e 2, 30167 Hannnover, Germany}
\affil[4]{\small Max-Planck-Institut for Gravitational Physics (Albert Einstein Institute) and \authorcr Institute for Gravitational Physics, \authorcr Leibniz Universit\"at Hannover, 30167 Hannover, Germany}
\affil[5]{\small Department of Physics, Technical University of Denmark,\authorcr
Fysikvej, 2800 Kgs.\,Lyngby, Denmark}
\affil[6]{\small Department of Physics, Graduate School of Science, University of Tokyo,\authorcr
7-3-1 Hongo, Bunkyo-ku, Tokyo, Japan, 113-0033}

\begin{document}

\maketitle

\begin{abstract}
An information reconciliation method for continuous-variable quantum key distribution with Gaussian modulation that is based on non-binary low-density parity-check (LDPC) codes is presented. Sets of regular and irregular LDPC codes with different code rates over the Galois fields $\GF(8)$, $\GF(16)$, $\GF(32)$, and $\GF(64)$ have been constructed. We have performed simulations to analyze the efficiency and the frame error rate using the sum-product algorithm.
The proposed method achieves an efficiency between $0.94$ and $0.98$ if the signal-to-noise ratio is between $4$ dB and $24$ dB.
\end{abstract}


\section{Introduction}

Quantum key distribution (QKD) \cite{Bennett_84, Gisin_02} allows two remote parties to establish an information-theoretically secure key. However, due to noise in the quantum channel and imperfections in quantum state preparation and measurement, errors (discrepancies) in the raw keys of the parties are unavoidable and have to be corrected. Consequently, a certain amount of information about the raw keys needs to be disclosed during an information reconciliation (error correction) process. Since the amount of disclosed information reduces the key rate, highly efficient information reconciliation methods are important for QKD systems. 

In typical discrete-variable (DV) QKD protocols as, e.g., the Bennett-Brassard 1984 (BB84) protocol \cite{Bennett_84}, the raw key is bit-wise encoded for the quantum communication. Hence, standard binary codes, which are highly efficient and have a large throughput, can be used for information reconciliation. Examples for such codes are, for instance, Cascade \cite{Pedersen_15, Martinez_15} or (rate-adapted) low-density parity-check (LDPC) codes \cite{Elkouss_11, Martinez_12, Martinez_13}. 

The situation is significantly different for continuous-variable (CV) QKD protocols in which quantum communication with a continuous encoding is used (see, e.g.,~\cite{weedbrook2012}). In order to generate the raw key, the continuous signals are then analog-to-digital converted (ADC) to obtain discrete values (symbols). The better the channel quality is (i.e., the larger the signal-to-noise ratio), the larger is the number of different values that can be distinguished. This number can be much greater than two and then the problem of efficiently reconciling raw keys is more challenging than in DV QKD.

Continuously modulated CV QKD protocols are usually based on Gaussian states that are normally distributed in the phase space. The quantization levels of the aforementioned ADC influence the distribution of the resulting raw key symbols. Although our reconciliation scheme would tolerate general quantization levels, in the following we consider only equidistant levels that are compatible with the security proof against general attacks in~\cite{Furrer_12,Furrer_14}. This has the consequence that the key symbols are not uniformly distributed. Thus, if each symbol is presented as a bit sequence, not all bit sequences are equally probable and the bits are not statistically independent.


Taking this into consideration, we detail in this work a reconciliation method that does not operate on the bit level but directly operates on the symbol level. This method, which we originally proposed for  the CV QKD protocol in \cite{Gehring_15}, is based on the belief propagation decoding of LDPC codes over Galois fields of the form $\GF(2^q)$ \cite{Davey_98, Declercq_07, Barnault_03}. We employ the sum-product algorithm, but use improved strategies for faster decoding that were recently proposed in~\cite{Voicila_10, Montorsi_12, Sayir_14}. Non-binary LDPC codes gained recently lots of interests due to several applications in different fields (see, e.g, Ref.~\cite{Arikan_15}).

We finally emphasize that any reconciliation method for QKD has to be compatible with the security proof. For instance, a requirement in most security proofs is that reconciliation has to be uni-directional. The case that Alice's raw key serves as reference, while Bob's raw key has to be reconciled is referred to as direct reconciliation. Alternatively, the term ``reverse reconciliation'' is used when Bob's raw key serves as reference. The reconciliation method that we propose here is applicable for both cases. 


The rest of the paper is organized as follows. In Section~\ref{sec:background}, we review previous approaches for information reconciliation for CV QKD. Section~\ref{sec:pre} provides the necessary details about the statistical properties of the signal generated by Gaussian modulated CV QKD protocols, and discusses the quantization of the signal (i.e., the analog-to-digital conversion). In Section~\ref{sec:method}, we describe the details of our reconciliation protocol. The performance of the codes is analyzed in Section~\ref{sec:results} using comprehensive simulations. Finally, we compare in Section~\ref{sec:discussion} the efficiency of our information reconciliation protocol with previously published methods.

\section{Related Work and our Contribution}
\label{sec:background}

Up to now different methods have been proposed for reconciling errors in CV QKD. Originally, an information reconciliation method referred to as sliced error correction (SEC) was proposed by Cardinal \textit{et al.} \cite{Cardinal_03, VanAssche_04, VanAssche_06}. It allows to reconcile the instances of two continuous correlated sources using binary error-correcting codes optimized for communications over the binary symmetric channel (BSC). In SEC, a set of $m$ slice (quantizing) functions and $m$ estimators are chosen to convert the outcome of each source into a binary sequence of length $m$. Each slice function, $s_i : \mathbb{R} \rightarrow \{0,1\}$ for $1 \le i \le m$, is used to map a continuous value to the $i$-th bit of the binary sequence. The corresponding $i$-th estimator $e_i$  is only used at the decoder side to guess the value of the transmitted $i$-th bit based on the received continuous value and the previously corrected slice bits from $1$ to $i-1$, given the knowledge of the joint probability distribution (correlation) of both sources. A communication model with individual BSCs per slice can then be considered and bit frames for each slice are independently encoded using an information rate depending on the associated channel. 
The slices $1,\dots,m$ are decoded successively; each decoded slice produces side information that can be used in the decoding of the following slices.\footnote{The side information from a decoded slice can also be used to improve the decoding of previous slices.}
Note that the encoding of each frame can be tackled with common coding techniques, and although it was initially proposed for turbo codes, the method was later improved using binary LDPC and polar codes \cite{Jouguet_13, Jouguet_14}.

Later, standard coding techniques such as multilevel coding (MLC) and multistage decoding (MSD) were proposed for reconciling errors in the Gaussian wire-tap channel, and in particular for CV QKD. Similar to SEC, MLC uses a quantization into slices to map the problem to individual BSCs. But the main difference stems from an improved decoding process. In MSD the resulting extrinsic information after decoding in each channel is used as a-priori information for decoding in another channel, thus, it works iteratively on the whole set of channels. Note that when only one iteration is performed for each level, this method is equivalent to SEC. Both techniques, MLC and MSD, were originally proposed for CV QKD in~\cite{Bloch_06a, Bloch_06b, Bloch_06c} using LDPC codes for decoding and considerably improving the efficiency of SEC for high SNRs.

Other methods and techniques, such as multidimensional reconciliation \cite{Leverrier_08a, Leverrier_08b} or multi-edge LDPC codes \cite{Jouguet_11}, were recently proposed for reconciling errors in CV QKD. These are, however, mainly focused on improving the reconciliation efficiency for low SNRs.

While LDPC codes over alphabets with more than two elements have already been introduced in the classic work by Gallagher \cite{Gallagher_63}, Davey and MacKay first reported that non-binary LDPC codes can outperform their binary counterparts under the message-passing algorithm over the BSC and the binary input additive white Gaussian noise channel (BI-AWGNC) \cite{Davey_98}. This behavior is attributed to the fact that the non-binary graph contains in general much fewer cycles than the corresponding binary graph \cite{Richardson_08}. Motivated by this fact, non-binary LDPC codes have been used in~\cite{Kasai_10} to improve the efficiency of information reconciliation in DV QKD. 

In this work we introduce the usage of non-binary LDPC codes for information reconciliation in CV QKD with Gaussian modulation and observe that this method reaches higher efficiencies (up to 98\%) than the previous approaches.

\section{Statistical characterization of the source}
\label{sec:pre}

We consider CV QKD protocols in which Alice's and Bob's raw keys are obtained from continuous variables that follow a bivariate normal distribution.
In an entanglement based description of CV QKD, these continuous variables are generated if Alice and Bob measure quadrature correlation of an entangled two-mode squeezed state of light (see, e.g.~\cite{weedbrook2012} and references therein). Equivalently, this can also be realized by a prepare-and-measure (P\&M) protocol in which Alice sends a Gaussian modulated squeezed or coherent state to Bob who measures the $Q$ or/and $P$ quadrature. Since it is conceptually simpler, we illustrate our results along an entanglement based CV QKD protocol in which both Alice and Bob measure either the $Q$ or $P$ quadrature. But the same reasoning can be applied to other Gaussian modulated CV QKD protocols.\footnote{We note that our error reconciliation based on LDPC codes can also be adapted to discrete modulated CV QKD protocols.}

In all what follows, we assume that Bob reconciles his values to match Alice's raw key, that is, direct reconciliation. However, due to the symmetry of the problem reverse reconciliation can be treated completely analogous by simply swapping Alice's and Bob's role. In the following sections, we discuss the classical statistical model of the aforementioned CV QKD protocols.     

\subsection{Model for normal source distribution} \label{sec:CVsource}

We give first a stochastic description of Alice's and Bob's continuously distributed measurement outcomes. 
If Alice and Bob measure the same quadrature $Q$ or $P$ of a two-mode squeezed state, their measurement outcomes are correlated or, respectively, anticorrelated. 
We denote the random variables corresponding to the measurement results of Alice and Bob in both quadratures by $Q_A$, $P_A$, $Q_B$, and $P_B$, respectively.
We assume that Alice and Bob remove all measurement values where they have not measured the same quadratures. 
To simplify the notation we introduce a new pair of random variables $(X_A,X_B)$ to denote either $(Q_A,Q_B)$ or $(P_A,-P_B)$.
We denote by $E(X)$ the expectation value of a random variable $X$ and by $\Gaussian(\mu,\sigma^2)$ the univariate normal (Gaussian) distribution with mean $\mu$ and standard deviation $\sigma$.

The random variables $X_A$ and $X_B$ are jointly distributed according to a bivariate normal distribution. Moreover, the marginal expectation values of $X_A$ and $X_B$ are both zero. The probability density function (pdf) of $X_A$ and $X_B$ can thus be written as
\begin{multline}
	\label{eq:JointProb}
	p(X_A=x_A, X_B=x_B) = \left( 2\pi \sigA\sigB\sqrt{1-\rho^2} \right)^{-1} \\
	\times \exp\left[
		-\frac{1}{1-\rho^2} \left( \frac{x_A^2}{2\sigA^2} + \frac{x_B^2}{2\sigB^2} - \rho \frac{x_A x_B}{\sigA \sigB} \right)
	\right],
\end{multline}
\noindent where $\sigA$ and $\sigB$ are the standard deviations of $X_A$ and $X_B$, respectively, and
\begin{equation}
\rho = E(X_A X_B) / \sigA \sigB
\end{equation}
\noindent is the correlation coefficient of $X_A$ and $X_B$. The covariance matrix is given by
\begin{equation} 
	\label{eq:cov}
	\vec{\Sigma}(X_A,X_B) = \begin{pmatrix}
		\sigA^2 & \rho \sigA \sigB \\
		\rho \sigA \sigB & \sigB^2
	\end{pmatrix}.
\end{equation}


Since the goal is to reconcile $x_B$ with $x_A$, Bob needs to know the conditional pdf's $p(x_A|x_B)$ for all $x_B$. We assume that Alice and Bob have performed a channel estimation (i.e., state tomography) to estimate the covariance matrix in Eq.~(\ref{eq:cov}) up to a small statistical error. The conditional pdf can be calculated from Eq.~\eqref{eq:JointProb} using $p(x_A|x_B) = p(x_A,x_B) / p(x_B)$, and is given by
\begin{equation}\label{eq:pcond}
	p(X_A=x_A|X_B=x_B)\sim\mathcal N(\mu_{A|B}(x_B),\sigma_{A|B}^2) \, ,
\end{equation}
\noindent with conditional mean and variance
\begin{align}
	\label{eq:condmean}
	\mu_{A|B}(x_B) := E(X_A|X_B=x_B) &= x_B \frac{\sigA}{\sigB} \rho \, , \\
	\label{eq:condvar}
	\sigma_{A|B}^2 := V(X_A|X_B=x_B) &= \sigA^2 (1-\rho^2) \, .
\end{align}

Note that the conditional variance is independent of Bob's measurement result $x_B$.

\subsection{Differential entropy and mutual information of the source}

We calculate now 
the mutual information between both sources $X_A$ and $X_B$.
%
We need some basic identities \cite[Chap.~9]{Cover_91}. The differential entropy of a continuous random variable $X$ with pdf $p(X)$ is given by $h(X) = - \int p(x) \log p(x) dx$. This allows us to introduce the differential conditional entropy of $A$ given $B$ as
\begin{align}\label{eq:diffcond}
	h(X_A|X_B) = h(X_A,X_B)-h(X_B) \, ,
\end{align}
\noindent and the mutual information between $X_A$ and $X_B$ as
\begin{align}\label{eq:mutinf}
I(X_A;X_B) = h(X_A) - h(X_A|X_B) \, .
\end{align}

The differential entropy of a univariate normal distribution with variance $\sigma^2$ is given by $h(X) = {1}/{2} \log_2 2\pi e \sigma^2$ and of a bivariate normal distribution with covariance matrix $\vec{\Sigma}$ by $h(X_A,X_B) ={1}/{2}\log_2\left((2\pi e)^2 \det \vec{\Sigma}\right)$. Hence, the mutual information of a bivariate normal distribution with covariance matrix given in Eq.~\eqref{eq:cov} can easily be computed as
\begin{align}\label{eq:mut-inf}
	I(X_A;X_B) = -\frac{1}{2}\log_2(1-\rho^2).
\end{align}

In accordance with the P\&M description of the protocol, we can think of $X_B$ as obtained by sending a Gaussian distributed variable $X_A$ with variance $\sigma_A^2$ through an additive white Gaussian noise channel (AWGNC). If the added noise variance of the AWGNC is $\sigma_N^2$, the mutual information between $X_A$ and $X_B$ is then given by
\begin{align}
	I(X_A;X_B) &= \frac{1}{2}\log_2\left(1 + \mathrm{SNR}\right),
\end{align}
\noindent where the signal-to-noise ratio is defined as $\text{SNR}=\sigma_A^2/\sigma_N^2$. This establishes a relation between the correlation coefficient $\rho$ and the SNR via 
\begin{align}\label{eq:SNR-rho}
	\mathrm{SNR} = \frac{\rho^2}{1-\rho^2} \, .
\end{align}

We finally emphasize that the mutual information only depends on $\rho$, but not on the marginal variances $\sigma_A$ and $\sigma_B$. This is clear since a rescaling of the outcomes $X_A$ and $X_B$ should not change the information between $X_A$ and $X_B$. It is thus convenient to work from the beginning with rescaled variables $Y_A$ and $Y_B$ such that the variance of both are $1$:
\begin{align} \label{eq:scaling}
	Y_A = \frac{X_A}{\sigA}, \qquad Y_B = \frac{X_B}{\sigB}.
\end{align}

Indeed, after the transformation we obtain for the marginal distributions of the scaled measurement outcomes $Y_A\sim \Gaussian(0,1)$, $Y_B\sim \Gaussian(0,1)$, and for the covariance matrix
\begin{equation} 
	\label{eq:rescaledcov}
	\vec{\Sigma}(Y_A,Y_B) = \begin{pmatrix}
		1 & \rho  \\
		\rho  & 1
	\end{pmatrix}.
\end{equation}

Equations \eqref{eq:pcond}--\eqref{eq:condvar} simplify to
\begin{equation}
\label{eq:scaledCond}
	p(Y_A=y_A|Y_B=y_B)\sim\mathcal N(y_B \rho, 1-\rho^2) \, .
\end{equation}

\subsection{Quantization of the continuous source}
\label{sec:Quantization}

In order to form the raw keys, the measurement results have to be quantized to obtain elements in a finite key alphabet $\alphabet=\{0,1,\cdots,2^p -1\}$. Such a quantization is determined by a partition of $\mathbb R$ into intervals, i.e. $\cP = \{I_k\}_{k\in\alphabet}$ (such that $\mathbb R = \bigcup_k I_k$ and $I_k\cap I_l = \emptyset$ for all $k\neq l$). Given a partition $\cP$, we define the quantization function $\quant_\cP$ by
\begin{align}
	\quant_\cP(y)=k \quad \text{if} \quad  y\in I_k \, .
\end{align}

In the following we consider specific partitions that are compatible with the security proof in~\cite{Furrer_12}. However, we emphasize that our results can be adapted to different partitions, which can be favorable if no requirements from the security proof have to be satisfied. The requirements on the partitions in~\cite{Furrer_12} are that a finite range $[-\alpha,\alpha)$ is divided into intervals of constant size $\delta>0$. Here the cut-off parameter $\alpha$ is chosen such that events $|Y_A|\geq  \alpha$ appear only with negligible probability. In order to complete the partition, outcomes in $[\alpha,\infty)$ and $(-\infty,-\alpha]$ are assigned to the corresponding adjacent intervals in $[-\alpha,\alpha)$.  More explicitly, this means that $I_k:=[a_k, b_k)$ with
\begin{align}
	a_k& = \begin{cases}
		-\infty & \text{if } k=0,\\
		-\alpha+k\delta \phantom{(k+1)\delta} & \text{if } k\in \alphabet\setminus\{0\},
	\end{cases}
\intertext{and}
	b_k& = \begin{cases}
		-\alpha+(k+1) \delta\phantom{k\delta} & \text{if } k\in \alphabet\setminus\{2^p-1\},\\
		\infty & \text{if } k=2^p-1 \, .
	\end{cases}
\end{align}


In the following, we only consider quantization maps with the above specified quantization characterized by $\alpha$ and $\delta$, and simply denote them by $\quant$ without specifying the partition. Moreover, for such a quantization map $\quant$, we will denote the discrete random variable obtained by applying it to a continuous variable $Y$ by $Z=\quant(Y)$. 

\subsection{Conditional quantized probability distribution and its mutual information}

Let $\quant$ denote a quantization map with fixed $\alpha$ and $\delta$. To reconcile a key symbol Bob does not need to know $Y_A$, but only the corresponding key symbol $Z_A=\quant(Y_A)$ that Alice has derived from $Y_A$.  Note, that we work in the following with the normalized variables $Y_A$ and $Y_B$ as defined in Eq.~\eqref{eq:scaling}.
Hence, for the decoding algorithm it is important to know the conditional probability of Alice's quantized variable $Z_A=\quant(Y_A)$ conditioned on $Y_B$. It is easy to calculate that for a bivariate normal source with covariance matrix given in Eq.~\eqref{eq:rescaledcov}, the probability that $Z_A=k$ (i.e., Alice's measurement $y_A$ is in the interval $I_k$) conditioned that Bob measures $y_B$ is given by%
\footnote{The cumulative distribution function $F_Y(y) = p(Y \le y)$ of the normal distribution $\Gaussian(\mu,\sigma^2)$ is $F(y;\mu,\sigma)= \Phi\left(\frac{y-\mu}{\sigma}\right) = \frac{1}{2} \left[ 1 + \erf \left( \frac{y-\mu}{\sqrt{2 \sigma^2}} \right) \right]$.}
\begin{align}
	p(Z_A& = k|Y_B=y_b) = p(Y_A \in I_k | Y_B=y_B) \label{eq:condPYA_in_Ik} \\
	& = \int_{I_k} p(Y_A=y_A|Y_B=y_B) \, d y_A  \nonumber  \\ 
	& = \frac{1}{2}\erf\left(\frac{b_k-y_B\rho}{\sqrt{2(1-\rho^2)}}\right) -\frac{1}{2}\erf\left(\frac{a_k-y_B\rho}{\sqrt{2(1-\rho^2)}}\right). \nonumber
\end{align}

To calculate the efficiency of a code, we first need to calculate the mutual information between $Z_A=\quant(Y_A)$ and $Y_B$. It is convenient to approximate the discrete entropic measures by their differential counterparts, which is well justified for quantizations considered in this article. The Shannon entropy of Alice's quantized source is given by $H(Z_A) = - \sum_k p(Z_A=k) \log_2 p(Z_A=k)$. For sufficiently small $\delta$ and sufficiently large $\alpha$, the entropy can be approximated as $H(\quant(Y_A)) \approx h(Y_A) - \log_2\delta$ (see, e.g.,~\cite[Chapt. 9]{Cover_91}). This also holds for the conditional entropy, that is, $H(\quant(Y_A)| Y_B) \approx h(Y_A| Y_B) - \log_2\delta$. Hence, it follows according to the definition of the mutual information (see Eq.~\eqref{eq:mut-inf}) that for appropriate $\delta$ and $\alpha$
\begin{equation}
I(\quant(Y_A); Y_B) \approx I(Y_A;Y_B)\,  , 
\end{equation}
\noindent where equality is obtained for $\alpha\rightarrow \infty$ and $\delta\rightarrow 0$. For the sake of completeness, we note that this even holds for the mutual information between Alice's and Bob's quantized variables:
\begin{equation}
I(\quant(Y_A); \quant(Y_B)) \approx I(Y_A;Y_B)\,  , 
\end{equation}
\noindent and equality holds for $\alpha\rightarrow \infty$ and $\delta\rightarrow 0$.

\section{Reconciliation Protocol}
\label{sec:method}

After the discussion of the statistical properties of the input source, we are ready to present our reconciliation protocol. We start with some preliminaries about reconciliation protocols in general and non-binary codes in particular.   

\subsection{Efficiency of a reconciliation protocol} \label{sec:efficiency}

The process of removing discrepancies from correlated strings is equivalent to source coding with side information at the decoder, also known as Slepian-Wolf coding \cite{Slepian_73}. In the asymptotic scenario of independent and identically distributed sources described by random variables $X$ and $Y$, the minimal bit rate at which this task can be achieved is given by $H(X|Y)$. Hence, the asymptotic optimal \emph{source coding} rate in our situation is simply given by
\begin{equation}\label{eq:sourcerate}
	R^{\text{source}}_{\text{opt}} = H(\quant(Y_A)|Y_B) \, . 
\end{equation}  
If the binary logarithm is used to calculate the conditional entropy in Eq.~\eqref{eq:sourcerate} the unit on both sides is bits/symbol and thus the numerical value can be larger than one.

In practical reconciliation algorithms the required source coding rate $R^{\text{source}}$ is generally larger than $R^{\text{source}}_{\text{opt}}$, because the number of samples (frame size) is finite and the reconciliation algorithm may not be optimal. A refined analysis of the optimal reconciliation rate for finite frame sizes has recently been given in~\cite{tomamichel2014}. 
For QKD reconciliation protocols it is common to define the efficiency $\beta\leq 1$ by the fraction of the mutual information that the protocol achieves \cite{Jouguet_14}.
Hence, the efficiency is calculated as
%
%
\begin{equation}
\label{eq:efficiency}
\beta = \frac{H(\quant(Y_A)) - R^{\text{source}}}{I(Y_A;Y_B)}.
\end{equation}
The efficiency can be factored as
\begin{equation}
\label{eq:efficiencyprod}
\beta = \beta_{\quant} \beta_{\mathrm{code}},
\end{equation}
where the quantization efficiency is given by
\begin{equation}
\label{eq:efficiencyq}
\beta_{\quant} = \frac{I(\quant(Y_A);Y_B)}{I(Y_A;Y_B)},
\end{equation}
and the efficiency of the coding is given by
\begin{equation}
\label{eq:efficiencyc}
\beta_{\mathrm{code}} = \frac{H(\quant(Y_A)) - R^{\text{source}}}{I(\quant(Y_A);Y_B)}.
\end{equation}

\subsection{Non-binary LDPC codes}

Linear codes have been used for decades for the purpose of correcting bit errors due to e.g. noisy transmission channels.
A linear code can be specified by a so-called parity check (PC) matrix $\vec H$. The specific feature of a low-density parity-check (LDPC) code is the fact that it has a \emph{sparse} PC matrix. 
Codes that have the same number of non-zero entries in each row and column of their PC matrix are called regular codes, otherwise they are called irregular.

The set $\mathcal{C}$ of all codewords of any linear code is formed by the kernel of $\vec{H}$, i.e., $\mathcal{C}:=\{\vec{x}: \vec{x}\vec{H}^\intercal=\vec{0}\}$. Typically, $\vec{H}$ is a binary matrix, and the code is used to correct binary values.
However, here we will use non-binary LDPC codes with PC matrices formed by elements of finite fields to correct symbols.
For convenience and faster decoding\cite{Barnault_03}, we only consider finite fields of order $2^q$, i.e., $\GF(2^q)$, although this is not crucial for our approach. 

For details about the construction of the PC matrices used in this work we refer to Section~\ref{sec:results}.

\subsection{Description of the non-binary reconciliation protocol}
\label{sec:decoding}

In this section we present our information reconciliation method. It is convenient to divide it into three different phases. In the first phase the measurement outcomes are collected, scaled and quantized as discussed in Section~\ref{sec:pre}. In the second phase the quantized outcomes are divided into least and most significant bits and the least significant bits are directly transmitted. In the third phase a non-binary LDPC code is used to reconcile the remaining most significant bits of each symbol. We present in the following the details of each phase. 

\subsubsection{Data representation}

Since we use a linear block code, Alice and Bob have to collect their measurement outcomes in a buffer until the number of measurements reaches the block size $n$ of the linear code. So, every time these buffers contain $n$ values Alice and Bob each form a frame, $\measXA, \measXB$, consisting of $n$ measurement outcomes, i.e., $\measXA, \measXB \in \mathbb{R}^n$. Alice and Bob scale their frames $\measXA,\measXB$ as in Eq.~\eqref{eq:scaling} to obtain the frames $\measYA,\measYB \in \mathbb{R}^n$, respectively. As discussed in Section~\ref{sec:pre}, we can assume that $\measYA, \measYB$ are obtained by $n$ independent samples of random variables $Y_A$ and $Y_B$ that follow a normal bivariate distribution with covariance matrix $\vec{\Sigma}$, as defined in Eq.~\eqref{eq:rescaledcov}. 

Alice quantizes her frames $\measYA$ by using a quantization map $\quant$ as introduced in Section~\ref{sec:Quantization} with predetermined values $\alpha$ and $\delta$. We assume that $\alpha$ and $\delta$ are given protocol parameters that may depend on the security proof of the CV QKD protocol for which the reconciliation is used (see, e.g.,~\cite{Furrer_12}). We denote the quantized frames by $\measZA\in\alphabet^n$. For further processing, Alice 
represents each symbol $k \in \alphabet=\{0,1,\dots, 2^p-1\}$ with $p$ bits using the binary representation $k_{p-1}\dots k_0$ determined by the decomposition $k=\sum_{i=0}^{p-1} k_i2^i$.  In the following, we identify $k\in \alphabet$ with its binary representation. 

\subsubsection{Separation of strongly and weakly correlated bits and disclosure of weakly correlated bits}

The binary representation of each symbol $k$ is divided into a pair of two shorter binary strings: $k=(\hat k,\check k)$, such that $\hat k \in \hat\alphabet :=\{0,1\}^q$ holds the $q$ most significant bits $k_{p-1}\dots k_{p-q}$ and $\check k \in \check\alphabet :=\{0,1\}^d$ holds the remaining $d=p-q$ least significant bits $k_{d-1}\dots k_0$, and $\alphabet=\hat\alphabet \times \check\alphabet$. 
Accordingly, Alice splits her frame $\rawA$ into a frame consisting of the $q$ most significant bits of each symbol, $\rawmA \in {\hat\alphabet}^n$, and a frame consisting of the remaining bits of each symbol, $\rawlA\in {\check\alphabet}^n$. Alice and Bob choose the value $q$ such that $\rawmA$ and $\measYB$ are sufficiently correlated to allow for non-trivial error correction, while $\rawlA$ and $\measYB$ are so weakly correlated that reconciliation can be done efficiently by a full disclosure.\footnote{It is clear that the splitting into strongly and weakly correlated bit depends on the initial symbol distribution. Hence, this step has to be adapted if one considers different (e.g., non-Gaussian) symbol distributions.}
Consequently, Alice sends through a noiseless channel the frame consisting of the $d$ least significant bits, $\rawlA$, to Bob, who sets $\rawlB = \rawlA$.  

\begin{figure}[t!]
\centering
\includegraphics[width=0.8\linewidth]{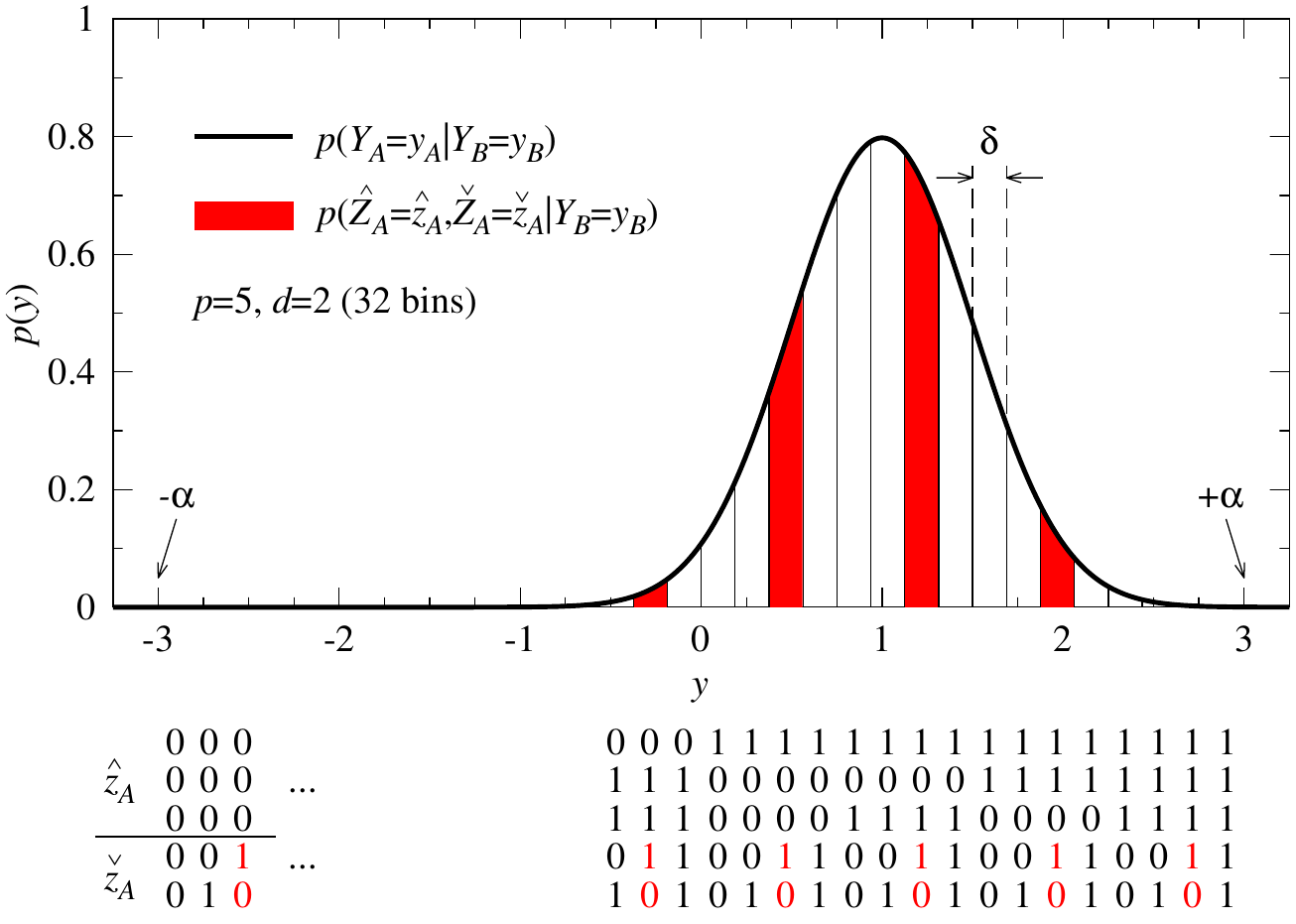}
\caption{Quantization of measurement results. Shown is the conditional probability as given in Eq.~\eqref{eq:scaledCond} (solid line)
for $\rho=\sqrt{3/4}$ and $y_B=\sqrt{4/3}$
	and its quantized version. For the latter we use a cutoff parameter $\alpha=3$ and divide the reconciliation interval in $32$ sub-intervals (bins) of width $\delta$. The bins are numbered with $p=5$ bits using the binary representation of 0 to 31. The area of each bin under the solid curve represents its occurence probability given by Eq.~\eqref{eq:condPYA_in_Ik}. We highlight the case where then the $d=2$ least significant bits have been disclosed as $\rawlA=(1,0)$ (marked in red color). The red areas correspond to the probabilities in the last row of Eq.~\eqref{eq:condPhatZA}.}
\label{fig:grid}
\end{figure}

The benefit of transmitting $\rawlA$, which is typically also performed in SEC \cite{VanAssche_06}, is that it helps to localize the symbols (i.e., it reduces the possible values for $y_A$ to the intervals that correspond to the filled areas in Fig.~\ref{fig:grid}) which leads to more accurate probabilities for the individual symbols in $\rawmA$ ($\rawmB$) and thus improves the efficiency of the next step. An example of this effect is depicted in Fig.~\ref{fig:grid}.
However, $d$ has to be chosen carefully as the least significant bits are transmitted directly, i.e., at a rate $R^\text{source}=1$. Therefore, to achieve a high efficiency, $d$ should be chosen such that $\rawlA$ and $Y_B$ are almost completely uncorrelated. Otherwise, Alice sends redundant information, which decreases the efficiency of the protocol.

\subsubsection{Reconciliation with non-binary LDPC code}

In the final step we use a non-binary LDPC code so that Bob can derive Alice's most significant bits $\rawmA$. 
Hence, as described in~\ref{sec:results}, Alice generates a suitable PC matrix $\vec{H}$ computes the syndrome $\rawmA \vec{H}^\intercal$ and sends it through a noiseless channel to Bob.\footnote{Note that the reconciliation efficiency depends on the code rate, which must be adapted depending on the correlation between $\rawmA$ and $\rawmB$ (see Section~\ref{sec:results}).} Then, Bob begins the decoding process by using an iterative belief propagation based algorithm that makes use of the syndrome value and the a-priori symbol probabilities for each element of the alphabet $\hat{\alphabet}$ for each symbol $\hat z_A$ in $\rawmA$ to derive $\rawmB$. The a-priori symbol probabilities are derived from $\rawlA$ and $\measYB$ using Bayes' rule:
\begin{align} \label{eq:condPhatZA}
	p(\hat Z_A &=\hat z_A|Y_B=y_b,\check Z_A = \check z_A) \nonumber \\
	& = \frac{p(\hat Z_A=\hat z_A,\check Z_A = \check z_A|Y_B=y_b)}%
	       {\sum_{\hat k\in \hat{\alphabet}}p(\hat Z_A=\hat k,\check Z_A = \check z_A|Y_B=y_b)} \\
	& = \frac{p(Z_A=(\hat z_A,\check z_A)|Y_B=y_b)}%
	       {\sum_{\hat k\in \hat{\alphabet}}p(Z_A=(\hat k,\check z_A)|Y_B=y_b)} \nonumber.
\end{align}

For Gaussian distributed symbols, the conditional probabilities in the last line of Eq.~\eqref{eq:condPhatZA} are calculated with the help of Eq.~\eqref{eq:condPYA_in_Ik}. In case that the decoder converges, $\rawmA$ and $\rawmB$ will coincide with high probability. Finally, Bob sets $\rawB := (\rawmB, \rawlB)$, using $\rawlB$ from the previous step.

We emphasize that the proposed non-binary reconciliation method applies also for sources with different statistical properties as long as the conditional probabilities in Eq.~\eqref{eq:condPhatZA} are available.

\newcommand{\Rsr}{R^{\textrm{source}}}
The source coding rate $\Rsr$ of this reconciliation protocol is given by the sum of the rates of the two steps which determine $\rawlB$ and $\rawmB$, respectively, i.e.,
\begin{equation}
\Rsr = 1\times d + R^\text{source}_{\text{LDPC}} \times q = d + (1-R_{\text{LDPC}} )q = p- q R_{\text{LDPC}},
\end{equation}
where we used that the \emph{channel coding rate} $R_{\text{LDPC}} $ of the LDPC code is related to its source coding rate via $R_{\text{LDPC}} =1-R^\text{source}_{\text{LDPC}} $.
$\Rsr$ forms an upper bound for the leakage:
\begin{equation}
\label{eq:leakage}
\text{leak} \le \Rsr.
\end{equation}

The efficiency, Eq.~\eqref{eq:efficiency} is then given by 
\begin{equation}
\beta = \frac{H(\quant(Y_A)) - p + q R_{\text{LDPC}}}{I(Y_A;Y_B)}.
\end{equation}

\section{Results}
\label{sec:results}



We performed simulations to analyze the frame error rate (FER), i.e., the ratio of frames that cannot be successfully reconciled, and the efficiency of regular and irregular non-binary LDPC codes.  The frame pairs $(\measYA,\measYB)$ for our simulations are generated by $n$ independent samples from joint random variables $(Y_A,Y_B)$ that follow a bivariate normal distribution with zero means, $\mu_A = \mu_B = 0$, unit variances, $\sigA^2 = \sigB^2 = 1$, and correlation coefficient $\rho$ as defined in Eq.~(\ref{eq:rescaledcov}). 

This is achieved by generating two independent unit normals $Y_1\sim \Gaussian({0,1})$ and $Y_2\sim \Gaussian({0,1})$ and using the transformation
\begin{eqnarray}
Y_A &=& Y_1, \\
Y_B &=& \rho Y_1 + \sqrt{1-\rho^2} Y_2.
\end{eqnarray}


We constructed ultra-sparse regular LDPC codes (with variable node degree $d_v=2$) and irregular LDPC codes over $\GF(8)$, $\GF(16)$, $\GF(32)$, and $\GF(64)$. Note, that in the following we use the symbol $R$ (instead of $R_{\text{LDPC}}$) to denote the channel code rate of LDPC codes.
The variable node degree distributions of the irregular LDPC codes were optimized using a differential evolution algorithm as described in~\cite{Shokrollahi_00}. The variable node degree distributions for $\GF(16)$ and $R=0.85$, $\GF(32)$ and $R=0.9$, and $\GF(64)$ and $R=0.9$, respectively, are given in Table~\ref{tab:2} of Appendix~\ref{sec:polynomials}. PC matrices for regular and irregular non-binary LDPC codes were then constructed using the progressive edge-growth algorithm described in~\cite{Hu_05}. Accordingly, we first constructed a binary PC matrix and then replaced every non-zero entry with a random symbol chosen uniformly from $\{1,2, \dots, 2^q-1\}$.

\begin{figure}[ht!]
\centering
\includegraphics[width=0.8\linewidth]{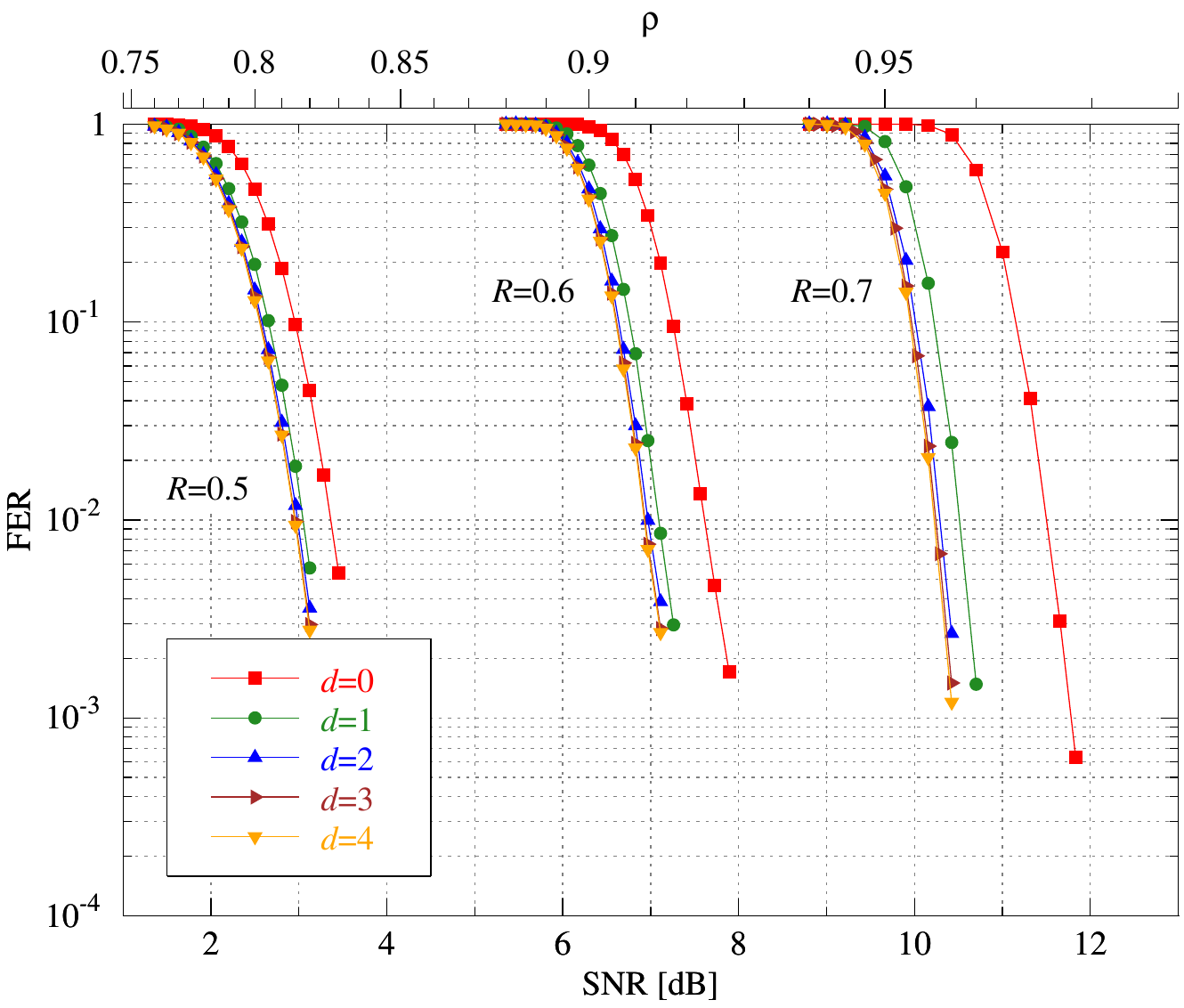}
\caption{Frame error rates of non-binary LDPC decoding over $\GF(32)$ for frame length $n=10^3$, cutoff parameter $\alpha=8$, and different code rates $R$, shown as a function of SNR (bottom axis) and $\rho$ (top axis).}
\label{fig:1}
\end{figure}

Non-binary LDPC decoding over $\GF(2^q)$ is performed using a sum-product (belief propagation based) algorithm. Given that codes are considered over a Galois field of order $2^q$, decoding was optimized using the $q$-dimensional Hadamard transform as proposed in~\cite{Barnault_03, Declercq_07}. The computational complexity per decoded symbol of this decoder is $\mathcal{O}(q 2^q)$. After each decoding iteration the syndrome of the decoded frame is calculated and the algorithm stops when the syndrome coincides with the one received from the other party (see Section~\ref{sec:method}) or when the maximum number of iterations is reached. When not explicitly stated, the maximum number of iterations in our simulations has always been 50.

\subsection{Performance}

\begin{figure}[ht!]
\centering
\includegraphics[width=0.8\linewidth]{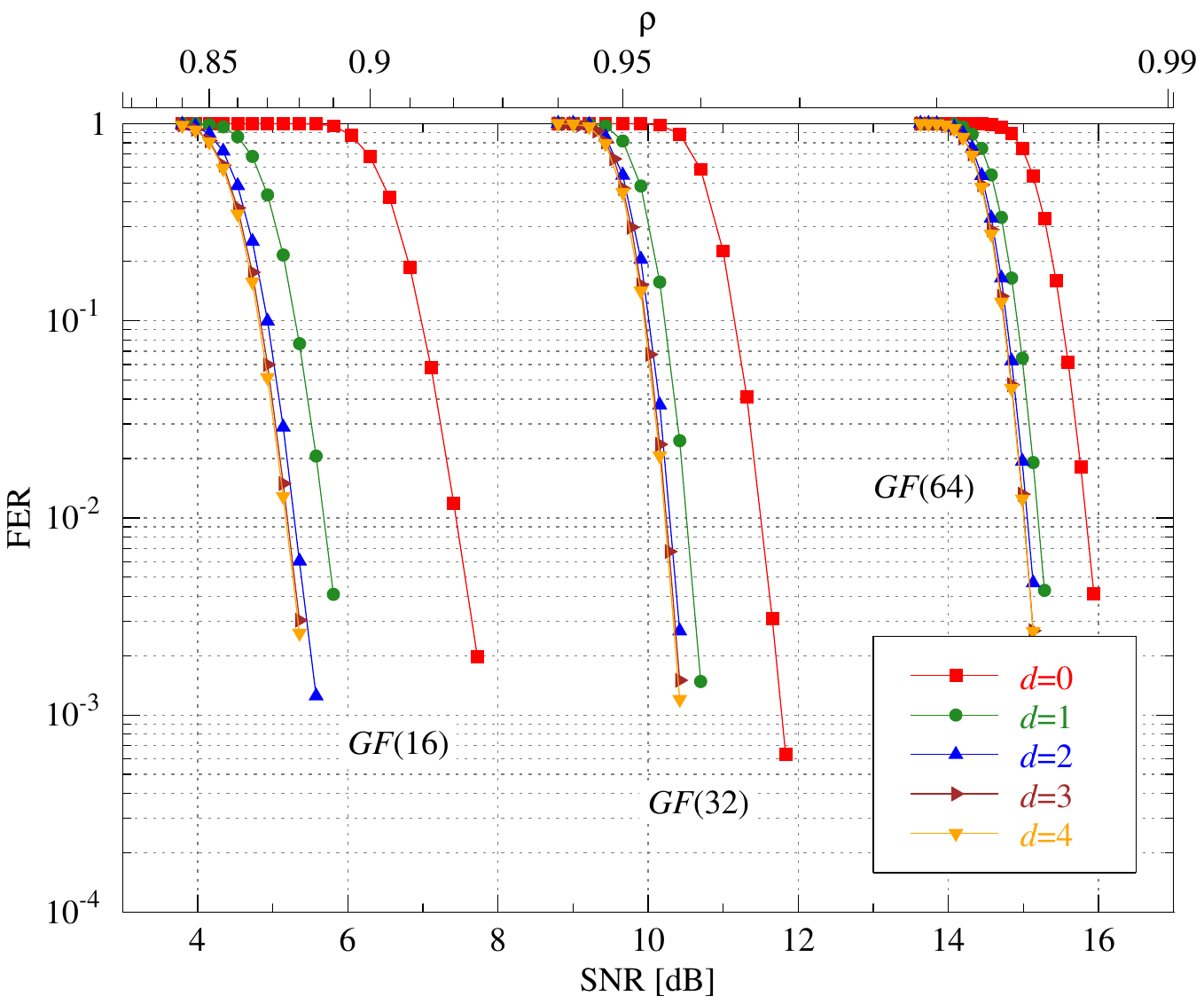}
\caption{Frame error rates of non-binary LDPC decoding over $\GF(16)$, $\GF(32)$, and $\GF(64)$, for frame length $n=10^3$,  cutoff parameter $\alpha=8$, and code rate $R=0.7$, shown as a function of SNR (bottom axis) and $\rho$ (top axis).}
\label{fig:2}
\end{figure}

\begin{figure}[ht!]
\centering
\includegraphics[width=0.8\linewidth]{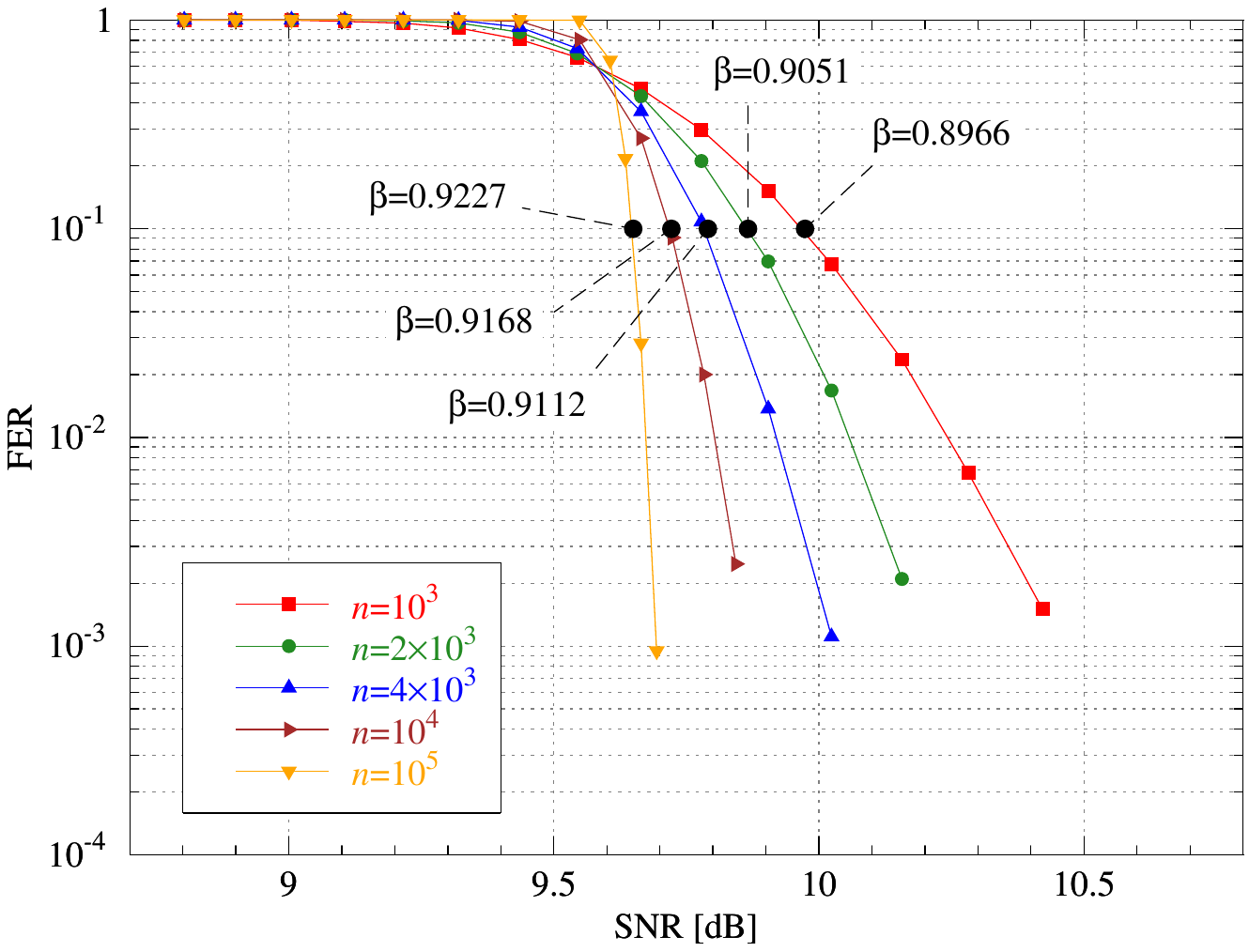}
\caption{Frame error rate (FER) for different frame lengths $n$ shown as a function of SNR. Parameters used: $q=5\sim$ $\GF(32)$, $R=0.7$, $\alpha=8$, $d=3$. For FER=10\% we denote the corresponding numerical values for the efficiency $\beta$.}
\label{fig:3}
\end{figure}

Figures~\ref{fig:1} to~\ref{fig:3} show the behavior of $(2,d_{c})$-regular non-binary LDPC codes for different numbers of sub-intervals of the reconciliation interval. The cutoff parameter is $\alpha=8$ for all curves shown. The FER is plotted as a function of the signal-to-noise ratio (SNR) in decibels (dB). In addition we show at the top X-axis the corresponding correlation coefficient $\rho$ that is related to the SNR via Eq.~\eqref{eq:SNR-rho}.

Fig.~\ref{fig:1} shows the FER of non-binary codes using a Galois field of order $32$, a short frame length of $n=10^3$ symbols, a cutoff parameter $\alpha=8$, and three different code rates. We observe that for code rates $R=0.5$, $R=0.6$, and $R=0.7$, the FER is monotonically decreasing in $d$ and saturates for $d = 3$.

Fig.~\ref{fig:2} also shows the FER for different numbers of sub-intervals of the reconciliation interval, but now we compare non-binary LDPC decoding over three different Galois fields $\GF(16)$, $\GF(32)$, and $\GF(64)$ for a fixed code rate $R=0.7$. As before, simulations were performed using regular non-binary LDPC codes with a frame length of $n=10^3$ symbols and $\alpha=8$. We observe the same monotonous and saturating behavior for the FER with increasing $d$ as in Fig.~\ref{fig:1}.
Although Fig.~\ref{fig:2} shows only the code rate $R=0.7$ we have confirmed this behavior for each Galois field for several code rates. The value $d=3$ has been empirically shown to be near optimal for all studied cases, even for different frame lengths and cutoff parameters. We conclude that $d=3$ is large enough to achieve near optimal frame error rate, and therefore, in the following we use $d=3$ to compute the frame error rate and reconciliation efficiency.

Fig.~\ref{fig:3} shows how the FER decreases with increasing frame length. Simulations were carried out using regular non-binary LDPC codes and decoding over $\GF(32)$ with the following parameters: code rate $R=0.7$, cutoff parameter $\alpha=8$, and number of least significant bits disclosed per symbol, $d=3$. The FER was computed and compared for five different frame lengths: $n=10^3$ symbols (red curve), $n=2\times 10^3$ (green), $n=4\times 10^3$ (blue), $n=10^4$ (brown), and $n=10^5$ (orange). In addition, the reconciliation efficiency $\beta$, cf.~Eq.~(\ref{eq:efficiency}), at a FER value of $10^{-1}$ (i.e., a success rate of 90\%) (solid black dots) is denoted for all frame lengths considered in the figure.
As shown, the efficiency increases with increasing frame length. Note also, that as expected, the increase of the efficiency is much larger when the frame length changes from $n=10^3$ to $n=10^4$ than the increase of the efficiency when going from $n=10^4$ to $n=10^5$.

\subsection{Reconciliation efficiency}

In the following we study the reconciliation efficiency $\beta$ of the proposed method as defined in Eq.~(\ref{eq:efficiency}) in more detail. Note that the efficiency of a code is calculated for a constant FER. Here, we considered a relatively high FER value of $10^{-1}$ in order to be able to compare our results with the literature \cite{Jouguet_11, Jouguet_13}.

\begin{figure}[ht!]
\centering
\includegraphics[width=0.8\linewidth]{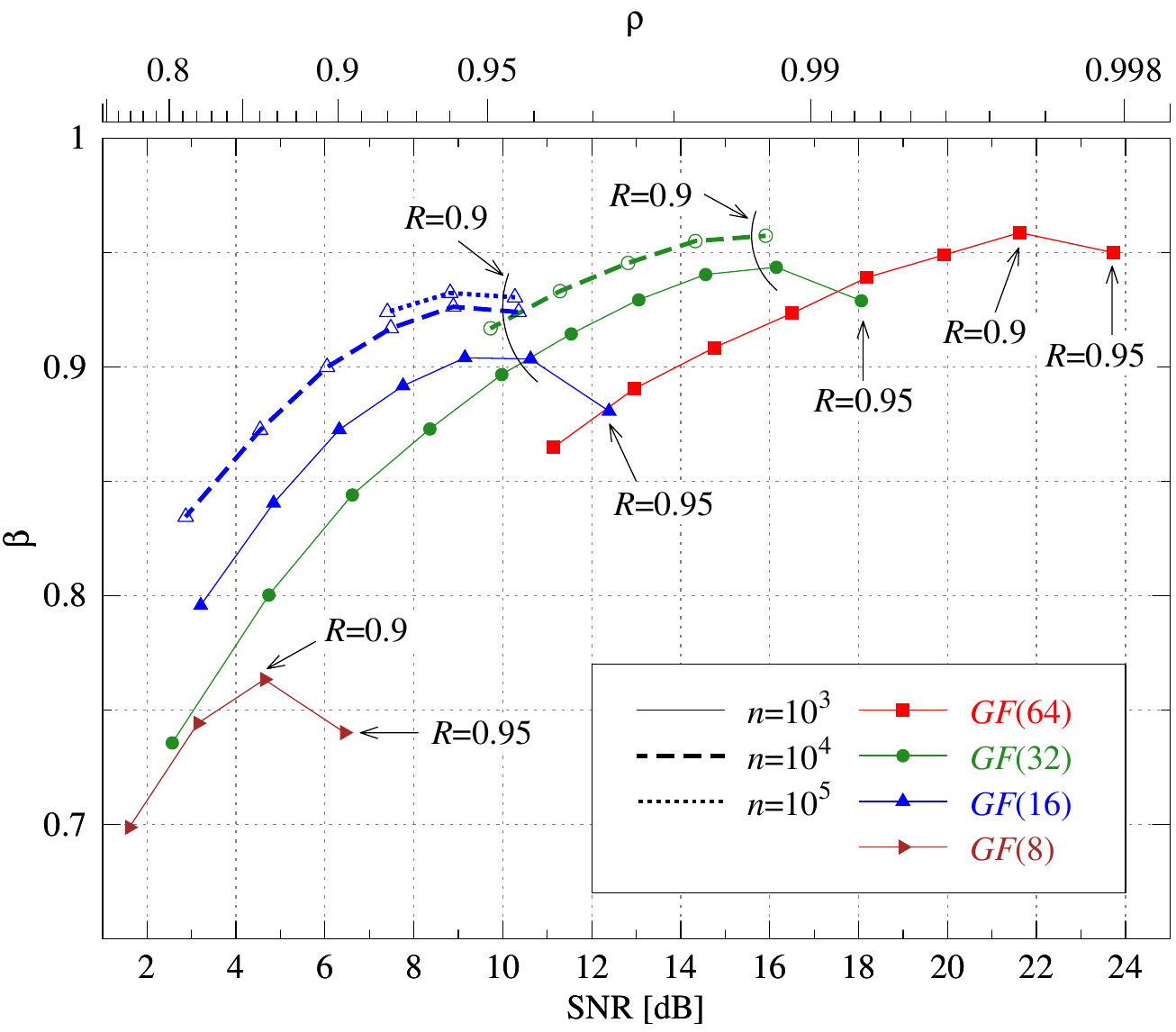}
\caption{Reconciliation efficiency of non-binary LDPC decoding over different Galois fields, using several frame lengths $n$ and code rates $R$. Note that for each line the code rates of two consecutive points differ by $0.05$. Common parameters: $d=3$, $\alpha=8$, and FER=$10^{-1}$. The latter parameter means that at each point a sequence cannot be reconciled in 10\% of cases.}
\label{fig:4}
\end{figure}

Fig.~\ref{fig:4} shows the reconciliation efficiency $\beta$ as a function of the SNR for non-binary LDPC decoding over different Galois fields, $\GF(8)$ (brown curve), $\GF(16)$ (blue), $\GF(32)$ (green), and $\GF(64)$ (red) for $n=10^3$ symbols (solid line). In addition we plot the efficiency also for larger frame lengths, i.e., for $n=10^4$ symbols (dashed line) for $\GF(16)$ and $\GF(32)$, and for $n=10^5$ symbols for $\GF(16)$ (dotted line). Simulations were carried out using regular non-binary LDPC codes, $d=3$ for the number of disclosed bits per symbols, and the cutoff parameter $\alpha=8$. Efficiency was calculated in all the cases estimating the highest SNR for which a sequence can be reconciled with a FER of $10^{-1}$. Several code rates were used to empirically estimate the expected reconciliation efficiency for a wide range of SNRs. Therefore, each point in the curves corresponds to the efficiency computed using a particular code rate (some of them labeled in the figure). Note that the code rate of two consecutive points on each curve differs by $0.05$.

\begin{figure}[ht!]
\centering
\includegraphics[width=0.8\linewidth]{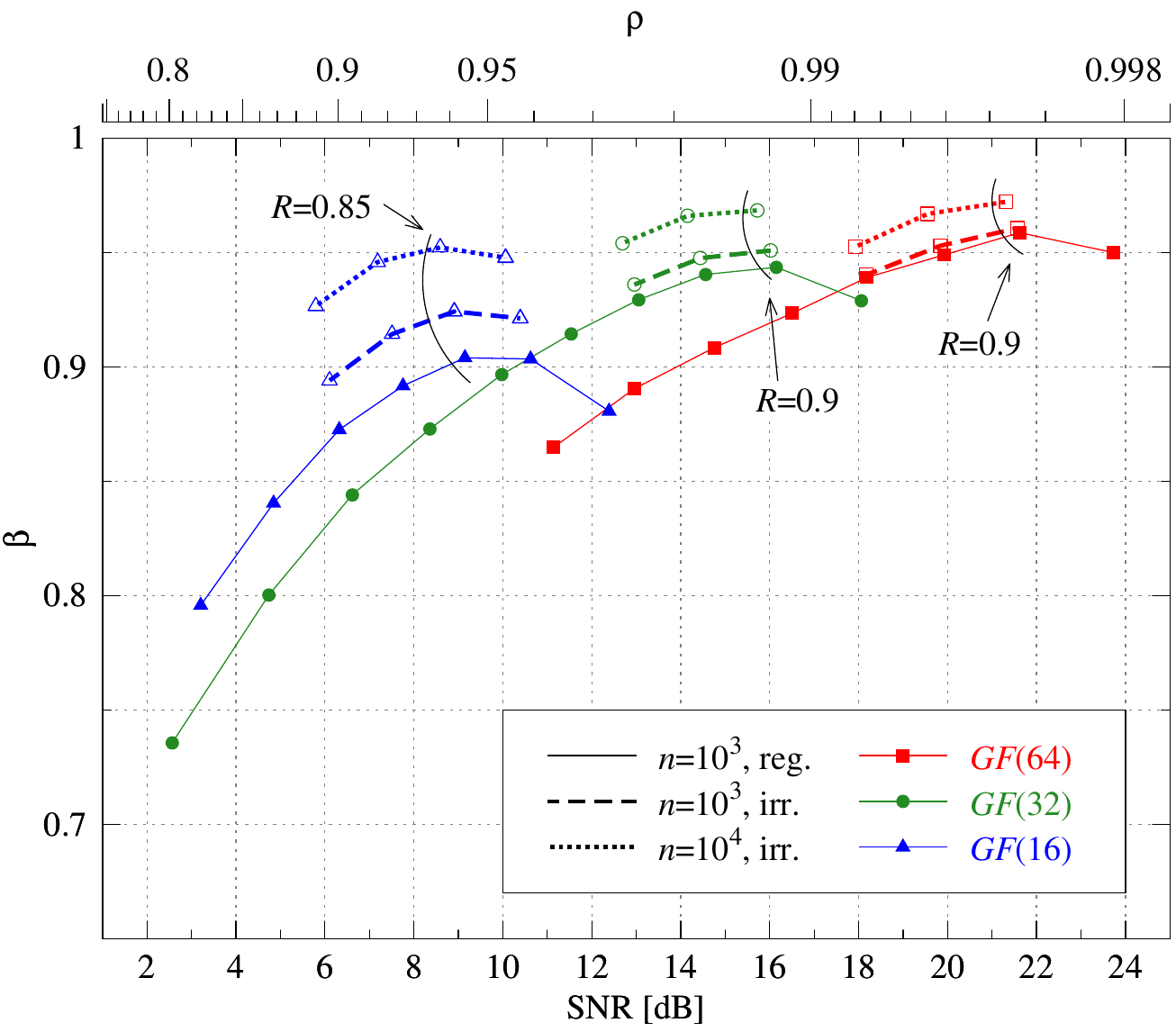}
\caption{Reconciliation efficiency with regular and irregular non-binary LDPC codes. As in Fig.~\ref{fig:4}, here it was considered non-binary LDPC decoding over different Galois fields, using several frame lengths $n$ and code rates $R$, with common parameters $d=3$, $\alpha=8$, and FER=$10^{-1}$.}
\label{fig:5}
\end{figure}

Fig.~\ref{fig:5} compares the results obtained with $(2,d_{c})$-regular codes of length $n=10^3$ (also shown in Fig.~\ref{fig:4}) with irregular codes of length $n=10^3$ and $n=10^4$. As previously, new simulations were computed for several code rates using the common parameters $d=3$, $\alpha=8$, and $\mathrm{FER}=10^{-1}$. Fig.~\ref{fig:5} shows how the reconciliation efficiency improves as the frame length increases and that irregular non-binary LDPC codes outperform regular non-binary LDPC codes particularly for lower Galois field orders. We observe that efficiency values above $0.95$ can be achieved for non-binary LDPC decoding over $\GF(16)$, $\GF(32)$ and $\GF(64)$ using irregular codes and frame lengths of $n=10^4$ symbols.

\begin{figure}[ht!]
\centering
\includegraphics[width=0.8\linewidth]{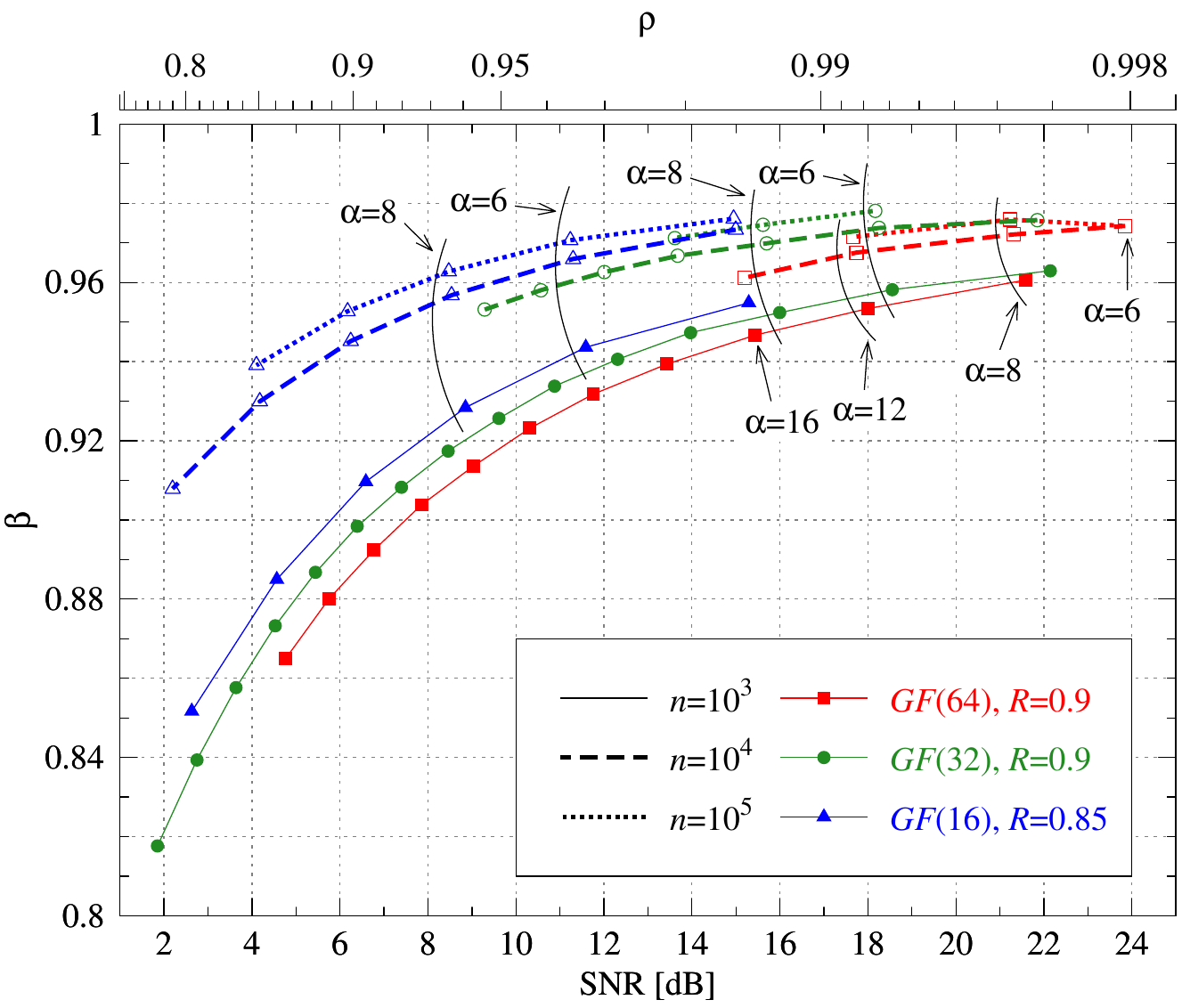}
\caption{Reconciliation efficiency of non-binary LDPC decoding over different Galois fields varying the cutoff parameter $\alpha$ for fixed-rate codes. Irregular non-binary LDPC codes and several frame lengths $n$ were considered, while $d=5$, $4$, or $3$ for decoding over $\GF(16)$, $\GF(32)$, or $\GF(64)$, respectively.}
\label{fig:6}
\end{figure}

Finally, Fig.~\ref{fig:6} shows the reconciliation efficiency as a function of the SNR for different cutoff parameters $\alpha$. Increasing values of $\alpha$ were considered for a constant code rate $R$. 
Fig.~\ref{fig:6} shows the efficiency of irregular non-binary LDPC codes for decoding over $\GF(16)$, $\GF(32)$, and $\GF(64)$, with code rates $R=0.85$, $R=0.9$, and $R=0.9$, respectively. In this case, the number of sub-intervals of the reconciliation interval remains constant at $2^9$, such that the number of disclosed bits differs for each Galois field, i.e., $d=5$, $4$, and $3$ for decoding over $\GF(16)$, $\GF(32)$, and $\GF(64)$, respectively. Some cutoff parameters are labeled in the figure. Note that the cutoff parameter of two consecutive points differ by $2$ (starting with $\alpha=4$) for those curves showing the decoding over $\GF(16)$ and $\GF(32)$, while consecutive points differ by $4$ for $\GF(64)$ except for the first point where $\alpha=6$ ($n=10^4$ and $10^5$). Finally, we conclude that the best efficiency is obtained by varying the cutoff parameter $\alpha$ of a fixed-rate code depending on the SNR. For a frame length of $n=10^4$ the efficiency is over $0.9$ in the range from $2$ to $24$~dB.

%

\section{Discussion}
\label{sec:discussion}

Here we propose the use of low-density parity-check codes over $\GF(2^q)$ for efficient information reconciliation in CV QKD. Although non-binary LDPC codes have a higher computational complexity (especially for large alphabets) than, for instance, binary LDPC codes, the benefit of using non-binary codes is potentially large \cite{Arikan_15}. In particular, there are several notable aspects of such codes that make this proposal interesting when compared with previous ones. Firstly, since a single communication channel is considered, only a single (non-binary) LDPC code needs to be optimized. This is in contrast to sliced approaches where the channel is divided into binary sub-channels. Secondly, all available information is used during the decoding process, that is, no information loss occurs through splitting of the data into slices. Consequently, as our results demonstrate, high efficiencies very close to unity can be achieved. Thirdly, although the amount of information disclosed in reconciliation is crucial, here we have shown that no rate-adaptive technique is needed to optimize the efficiency. Instead, by varying the width of the reconciliation interval (using a cutoff parameter $\alpha$) depending on the signal-to-noise ratio, sequences can be efficiently reconciled in a range of SNRs using only one fixed-rate code.

\begin{table*}[htbp!]
\caption{Efficiency values.}
\label{tab:1}
\centering
\begin{tabular}{|c|c||c|c|c|c|c|}
\hline
SNR (lin/dB) & $\rho$ & $\beta_{\mathrm{SEC}}$ & $\beta_{\mathrm{SEC}}$ & $\beta_{\mathrm{MSD}}$ & $\beta_{\mathrm{multi\textrm{-}dim}}$  & $\beta_{\mathrm{non\textrm{-}binary}}$ \\
\hline\hline
$0-1$ / up to $0$ & (0.707) & $60\%$ & $94.2\%$ & $79.4\%$ & $89\%$ & \\
$3$ /  $4.8$      & 0.866 & $79\%$ & $94.1\%$ & $88.7\%$ & $90\%$ & $94.3\%-95.2\%$ \\
$5$ /  $7.0$      & 0.913 &  --    & $94.4\%$ &   --     &  --    & $95.7\%-96.5\%$ \\
$7$ /  $8.5$      & 0.935 & $84\%$ &    --    & $90.9\%$ &  --    & $96.3\%-97.0\%$ \\
$15$ / $11.8$     & 0.968 & $92\%$ & $95.8\%$ & $92.2\%$ &  --    & $97.1\%-97.7\%$ \\
$31$ / $14.9$     & 0.984 &  --    &    --    &   --     &  --    & $97.6\%-98.2\%$ \\
\hline
$n$ (bits) & & $2 \times 10^5$ & $2^{20}\approx 10^6 $ & $2 \times 10^5$ & & $10^5$ (symbols) \\
\hline
Refs. & & \cite{Bloch_06a} & \cite{Jouguet_14} & \cite{Bloch_06a} & \cite{Leverrier_08a,Leverrier_08b} & this work \\
\hline
\end{tabular}
\end{table*}

Table~\ref{tab:1} summarizes (to the best knowledge of the authors) the best efficiency values for CV QKD reconciliation reported in the literature. In the table, three different information reconciliation techniques are compared with this work ($\beta_{\mathrm{non\textrm{-}binary}}$) for different ranges of SNRs: (1) sliced error correction ($\beta_{\mathrm{SEC}}$) originally proposed by Cardinal \textit{et al.} in~\cite{Cardinal_03, VanAssche_04} (using turbo codes) and later improved in~\cite{Jouguet_13, Jouguet_14} (using LDPC and polar codes), (2) multilevel coding and multistage decoding ($\beta_{\mathrm{MSD}}$) using LDPC codes \cite{Bloch_06a}, and (3) multidimensional reconciliation ($\beta_{\mathrm{multi\textrm{-}dim}}$) \cite{Leverrier_08a, Leverrier_08b, Jouguet_11}. The smaller value of $\beta_{\mathrm{non\textrm{-}binary}}$ is obtained for a maximum of 50 decoding iterations, while the larger value corresponds to simulations with a maximum number of 200 decoding iterations. As shown in Table~\ref{tab:1}, the proposed method improves all previously published values for the efficiency in the high SNR regime.

\section{Conclusions}

We presented an information reconciliation scheme for continuous-variable quantum key distribution that is based on non-binary LDPC codes. While we analyze its performance and efficiency for Gaussian distributed variables, the scheme is also well suited for other non-uniform symbol distributions. The reconciliation scheme is divided into two steps. First, the least significant bits of Alice's quantized variable -- typically $d=3$ in our simulations -- are disclosed. Then, the syndrome of a non-binary LDPC code is transmitted and used together with the information from the first step to reconcile the remaining significant bits of each measurement result. Using irregular LDPC codes over $\GF(2^q)$, this enabled us to achieve reconciliation efficiencies between 0.94 and 0.98 at a frame error rate of 10\% for signal-to-noise ratios between $4$ dB and $24$ dB. 


\section*{Acknowledgements
}
The authors thank Torsten Franz, Vitus H\"andchen, and Reinhard F. Werner for helpful discussions. This work has been partially supported by the Vienna Science and Technology Fund (WWTF) through project ICT10-067 (HiPANQ), and by the project Continuous Variables for Quantum Communications (CVQuCo), TEC2015-70406-R, funded by the Spanish Ministry of Economy and Competiveness. Fabian Furrer acknowledges support from Japan Society for the Promotion of Science (JSPS) by KAKENHI grant No. 12F02793.

\bibliographystyle{custom}
\bibliography{references}

\section*{Appendix}
\appendix

%
%
%
%
%
%
%
%
%
%

\section{Optimized Polynomials}
\label{sec:polynomials}

Table~\ref{tab:2} shows the generating polynomials that describe the ensemble of irregular LDPC codes used in Fig.~\ref{fig:6}.

\begin{table}[ht!]
\caption{Generating polynomials.}
\label{tab:2}
\centering
\begin{tabular}{|c|c|c|c|}
\hline
Coeff. & $\GF(16)$ & $\GF(32)$ & $\GF(64)$ \\
$\lambda(x)$ & $R=0.85$ & $R=0.9$  & $R=0.9$  \\
\hline\hline
$\lambda_{2}$  & $0.62755$ & $0.67173$ & $0.81173$ \\
$\lambda_{5}$  &           &           & $0.00710$ \\
$\lambda_{6}$  & $0.03896$ & $0.00164$ &           \\
$\lambda_{7}$  &           & $0.00481$ &           \\
$\lambda_{8}$  &           & $0.01342$ & $0.01004$ \\
$\lambda_{10}$ & $0.02497$ &           &           \\
$\lambda_{11}$ & $0.01158$ &           &           \\
$\lambda_{14}$ & $0.00598$ & $0.02081$ &           \\
$\lambda_{15}$ & $0.03557$ &           & $0.17113$ \\
$\lambda_{16}$ &           & $0.28759$ &           \\
$\lambda_{17}$ & $0.20497$ &           &           \\
$\lambda_{19}$ & $0.05042$ &           &           \\
\hline
\end{tabular}
\end{table}

%
%

\end{document}